# Design of High Performance MIPS Cryptography Processor Based on T-DES Algorithm

Kirat Pal Singh[1], Shivani Parmar[2]

[1]*Research Fellow,* [2]*Assistant Professor*

[1]*Academic and Consultancy Services Division,* [2] *ECE Department*

[1]*Centre for development of advanced computing (C-DAC),* [2]*GGS College*

[1, 2]*Mohali-160071, India*

## ABSTRACT

*The paper describes the design of high performance MIPS Cryptography processor based on triple data encryption standard. The organization of pipeline stages in such a way that pipeline can be clocked at high frequency. Encryption and Decryption blocks of triple data encryption standard (T-DES) crypto system and dependency among themselves are explained in detail with the help of block diagram. In order to increase the processor functionality and performance, especially for security applications we include three new 32-bit instructions LKLW, LKUW and CRYPT. The design has been synthesized at 40nm process technology targeting using Xilinx Virtex-6 device. The overall MIPS Crypto processor works at 209MHz.*

## 1. INTRODUCTION

Today's digital world, Cryptography is the art and science that deals with the principles and methods for keeping message secure. Encryption is emerging as a disintegrable part of all communication networks and information processing systems, involving transmission of data. Encryption is the transformation of plain data (known as plaintext) into inintengible data (known as cipher text) through an algorithm referred to as cipher. MIPS architecture employs a wide range of applications. The architecture remains the same for all MIPS based processors while the implementations may differ [1]. The proposed design has the feature of 32-bit asymmetric and symmetric cryptography system as a security application. There is a 16- bit RSA cryptography MIPS cryptosystem have been previously designed [2]. There are the small adjustments and minor improvement in the MIPS pipelined architecture design to protect data transmission over insecure medium using authenticating devices such as data encryption standard [DES], Triple-DES and advanced encryption standard [AES] [3]. These cryptographic devices use an identical key for the receiver side and sender side. Our design mainly includes the symmetric cryptosystem into MIPS pipeline stages. That is suitable to encrypt large amount data with high speed.

The MIPS is simply known as Millions of instructions per second and is one of the best RISC (Reduced Instruction Set Computer) processor ever designed. High speed MIPS processor possessed Pipeline architecture for speed up processing, increase the frequency and performance of the processor. A MIPS based RISC processor was described in [4]. It consist of basic five stages of pipelining that are Instruction Fetch, Instruction Decode, Instruction Execution, Memory access, write back. These five pipeline stages generate 5 clock cycles processing delay and several Hazard during the operation [2]. These pipelining Hazard are eliminates by inserting NOP (No Operation Performed) instruction which generate some delays for the proper execution of instruction [4]. The pipelining Hazards are of three type's data, structural and control hazard. These hazards are handled in the MIPS processor by the implementation of forwarding unit, Pre-fetching or Hazard detection unit, branch and jump prediction unit [2]. Forwarding unit is used for preventing data hazards which detects the dependencies and forward the required data from the running instruction to the dependent instructions [5]. Stall are occurred in the pipelined architecture when the consecutive instruction uses the same operand of the instruction and that require more clock cycles for execution and reduces performance. To overcome this situation, instruction pre-fetching unit is used which reduces the stalls and improve performance. The control hazard are occurs when a branch prediction is mistaken or in general, when the system has no mechanism for handling the control hazards [5]. The control hazard is handled by two mechanisms: Flush mechanism and Delayed jump mechanism. The branch and jump prediction unit uses these two mechanisms for preventing control hazards. The flush mechanism runs instruction after a branch and flushes the pipe after the misprediction [5]. Frequent flushing may increase the clock cycles and reduce performance. In the delayed jump mechanism, to handle the control hazard is to fill the pipe after the jump instruction with specific numbers of NOP's [5]. The branch and jump prediction unit placement in the pipelining architecture may affect the critical or longest path. To detecting the longest path and improving the hardware that resulting minimum clock period and is the standard method of increasing the performance of the processor.

To further speed up processor and minimize clock period, the design incorporates a high speed hybrid adder which employs both carry skip and carry select techniques with in the ALU unit to handle the additions.

This paper is organized as follows. The system architecture hardware design and implementation are explained in Section II. Instruction set of MIPS including new instructions in detail with corresponding diagrams shown in sub-sections. Hardware implementation design methodology is explained in section III. The experimental results of pipeline stages are shown in section IV. Simulation results of encrypted MIPS pipeline processor and their Verification & synthesis report are describes in sub sections. The conclusions of paper are described in section V.

## 2. SYSTEM ARCHITECTURE

The global architecture of encrypted and decrypted pipelined processor is shown in Fig.1 which contain 5 basic pipeline stages that are Instruction Fetch [IF], Instruction Decode [ID], Instruction Execution [EXE], Memory Access [MEM], Write Back [WB]. These pipeline stages operate concurrently, using synchronization signals: Clock and Reset. MIPS architecture employs three different Instruction format: R-Type Instruction, I-Type Instruction, and J-Type Instruction [3]. MIPS processor for encryption/decryption we just insert the cryptography module such as data encryption standard (DES), Triple data encryption standard (T-DES), advanced encryption standard (AES) etc. to the pipeline stage. Only single cryptographic module is used in same hardware implementation. In this design, we insert T-DES crypto core inside the instruction fetch stage and memory access stage for instructions fetching and data storing.

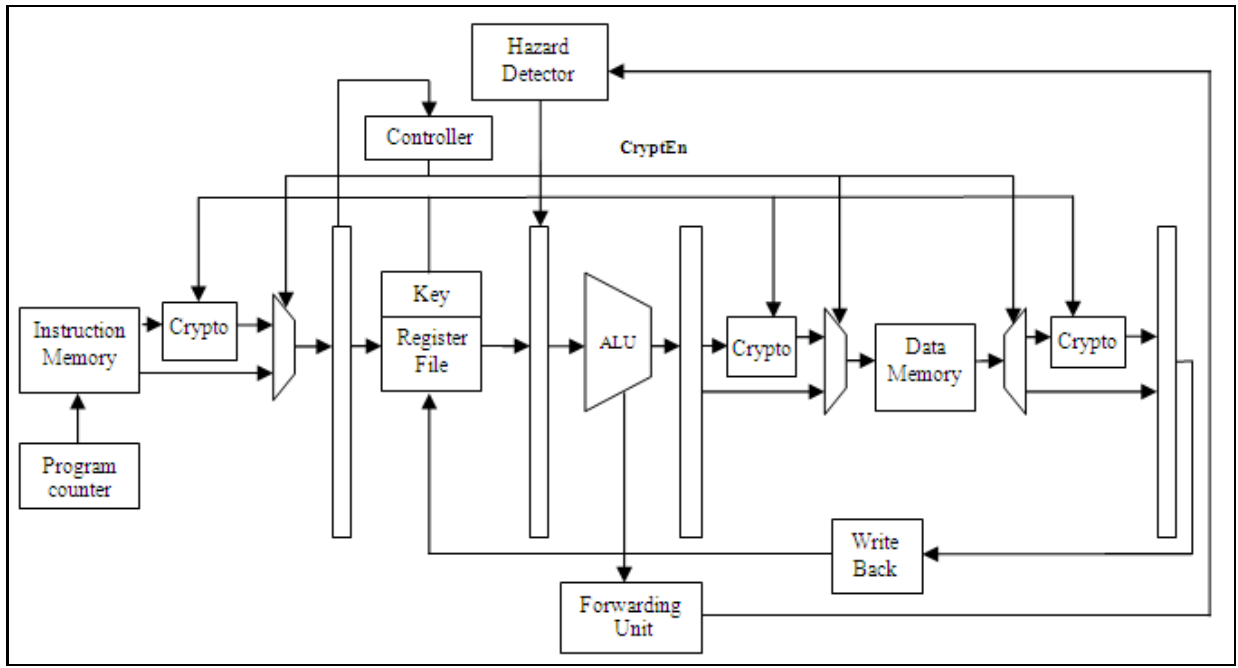

**Figure 1.Block diagram of Encrypted/Decrypted MIPS processor**

## 2.1 Encrypted MIPS Processor

The 32-bit Encrypted MIPS processor is generally based on MIPS architecture. The pipelined MIPS architecture is modified in such a way that it executes encrypted instruction. The function of Instruction fetch unit is to obtain an instruction from the instruction memory using the current value of PC and increment PC value for next instruction and placed that value to IF register. The instruction fetch unit of encrypted MIPS contains program counter (PC), instruction memory, T-DES decryption core and MUX. The instruction memory read address from PC and store instruction value at the particular address that points by the PC. Instruction memory sends encrypted instruction to MUX and decryption core. The decryption core give decrypted instructions and further send to the MUX and output of MUX is fed to the IF register. The MUX control signal comes from control unit. The instruction decode unit contain register file and key register. Key register store the key data of encryption/decryption core. Key address and key data comes from write back stage. Once the key data to be stored into register file it will remain same for all program instruction execution. The control unit provides various control signals to other stages. The execute unit executes the register file output data and perform the particular operation determined by the ALU. The ALU output data send to EXE register. The memory access unit contains T-DES encryption core, T-DES decryption core, data memory, MUX, DEMUX. The second register data from register file fed to the encryption core and also to MUX. Here the crypt signal enable/disable encryption operation. The read/write signal of data memory describes whether reading/writing operation is done. Output of data memory pass through DEMUX and its one output go to decryption core and other goes to MEM register. Here the unencrypted memory data and decrypted data temporarily store to MEM register. The MEM output fed to write back data MUX and according to control signal the output of MUX goes to register file.

## 2.2 Decrypted MIPS Processor

The 32-bit decrypted MIPS processor is generally based on MIPS architecture. The pipelined MIPS architecture is modified in such a way that it executes decrypted instruction. The instruction fetch unit of decrypted MIPS processor contains program counter (PC), instruction memory, T-DES encryption core and MUX. The instruction memory read address from PC and store instruction value at the particular address that points by the PC. Instruction memory sends decrypted instruction to both MUX and encryption core. The encryption core give encrypted instructions and further send to the MUX and output of MUX is fed to the IF register. The MUX control signal comes from control unit. The instruction decode unit contain register file and key register. Key register store the key data of encryption/decryption core. Key address and key data comes from write back stage. Once the key data to be stored into register file it will remain same for all program instruction execution. The control unit provides various control signals to other stages. The execute unit executes the register file output data and perform the particular operation determined by the ALU. The ALU output data send to EXE register. The memory access unit contains T-DES encryption core, T-DES decryption core, data memory, MUX, DEMUX. The second register data from register file fed to the decryption core and also to MUX. Here the crypt signal enable/disable decryption operation. The read/write signal of data memory describes whether reading/writing operation is done. Output of data memory pass through DEMUX and its one output go to encryption core and other goes to MEM register. Here the unencrypted memory data and encrypted data temporarily store to MEM register. The MEM output fed to write back data MUX and according to control signal the output of MUX goes to register file.

## 2.3 MIPS Instruction Set

The operational mode of the MIPS crypto processor is controlled by a RESET signal. When the RESET signal is at logic "0", the crypto processor is in the reset mode and the processing unit writes the memory and register contents using the 32-bit bidirectional data bus, 10-bit address bus, and four control signals. The keys are kept in the key registers of the register file of crypto processor that are available to other stages of processor.

The MIPS instruction set is straightforward like other RISC designs. MIPS are a load/store architecture, which means that only load and store instructions access memory. Other instructions can only operate on values in registers [6]. Generally, the MIPS instructions can be broken into three classes: the memory-reference instructions, the arithmetic- logical instructions, and the branch instructions. Also, there are three different instructions formats in MIPS architecture: R-Type instructions, I-Type instructions, and J-Type instructions as shown in Fig. 2.

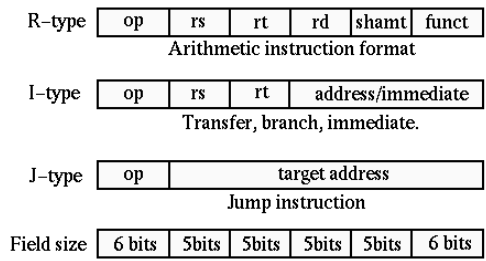

Figure 2.MIPS Instruction Type

The MIPS instruction field is described in Table 1. There are three more new instructions that support encrypted and decrypted operation. These instructions are load key upper word (LKUW), load key lower word (LKLW) and encryption mode (CRYPT). These instructions randomly used opcodes in the hardware implementation. LKLW and LKUW come under I-type instruction and variant of load word (LW). These two instruction need not to specify a destination address in the assembly code. CRYPT instruction comes under J-type instruction and instead of address, only single argument i.e. Boolean value is to be assigned. This indicates enable/disable encryption and decryption process. Any nonzero value enables the encryption/decryption process and zero value disables the encryption process.

Table 1. MIPS Instruction filed

| Field | Description |
|---|---|
| Op[31-26] | is a 6-bit operation code |
| RS[25-21] | is a 5-bit source register specifier |
| RT[20-16] | is a 5-bit target(source/destination) register or branch condition |
| Immediate[15-0] | is a 16-bit immediate, branch displacement or address displacement |
| Target[25-0] | is a 26-bit jump target address |
| RD[15-11] | is a 5-bit destination register specifier |
| Shamt[10-6] | is a 5-bit shift amount |
| Funct[5-0] | is a 6-bit function field |

## 3. IMPLEMENTATION METHODOLOGY

Current applications demand high speed processor for large amount of data transmission in real time. As compared to software alternatives, hardware implementation provides highly secure algorithms and fast solutions approaches for high performance applications. Software approaches could be a good choice but it has some limitations like low performance and speed. Main advantages of software are low cost and short time to market. But they are unacceptable in terms of high speed and performance specification. So that Hardware alternatives could be selected for implementing MIPS crypto processor architecture.

Hardware implementation supports both Field Programmable Gate Arrays (FPGAs) and Application Specific Integrated Circuits (ASIC) at high data rates. Such design has high performance but more time consuming and expensive as compared to software alternatives. The detailed comparison of hardware vs. software solutions for implementing the MIPS crypto processor architecture is shown in Table 2. Based on the comparison, hardware solution is a better choice in most of the cases because they have high performance. The main advantage of FPGA in hardware alternative, FPGA are low density and low area consumption.

Logic integration, size and density are the major drawbacks in ASIC but have higher performance than FPGA.

We use the T-DES Crypto core which supports both encryption and decryption. T-DES core has a 64-bit plaintext input, three 64-bit key input, start signal, encryption/decryption enable signal, and 64-bit cipher text output. The 32-bit encrypted processor pack two 32-bit MIPS instruction into a single 64-bit instruction block for DES encryption/decryption process and breakout each individual instruction into the hardware. In this processor, there is some unencrypted instruction stored in data memory as a zero-padded 64-bit word. The program counter increment by 8 instead of 4 due to loading of 64-bits instructions. Both data and instruction memory reads 64-bit instruction at a time.

Table 2. Hardware vs. software alternatives for crypto processor

| Parameters | Software | Hardware | |
|---|---|---|---|
| | | FPGA | ASIC |
| Performance | Low | Medium-High | Very High |
| Power consumption | Depends | Very high | Low |
| Logic integration | Low | Low | High |
| Tool cost | Low | Low | Low |
| Test development complexity | Very low | Very low | High |
| Density | High | Very low | High |
| Design efforts | Low-medium | Low-medium | High |
| Time consumed | Short | Short | High |
| Size | Small-medium | Small | Large |
| Memory | Fine | Fine | Fine |
| Flexibility | High | High | - |
| Time to market | Short | Short | High |
| Run time configuration | - | high | - |

Before fetching the encrypted instruction the key loaded from memory is to be done and there may be a dependency of instruction takes place that causes hazard. So to overcome this dependency we may use NOP instruction that deactivates the control signal of current instructions. The forwarding unit also modifying to give the load key instruction. The NOP's are sufficient to insert between load key and CRYPT instruction. This is explained by an example.
Example:-

```
addi $r1, $r0, 104
lklw 0($r1)
addi $r1, $r1, 8
lkuw 0($r1)
addi $r2, $r1, 8
lklw 0($r2)
addi $r2, $r2, 8
lkuw 0($r2)
addi $r3, $r0, 8
lklw 0($r3)
addi $r3, $r3, 8
lkuw 0($r3)
nop
nop
crypt 1
addi $r1, $r0, 7
add  $r2, $r0, $r0
addi $r3, $r0, 0
addi $r4, $r0, 0
Loop:  add $r5, $r2, $r2
add $r5, $r5, $r5
```

```
add  $r5, $r5, $r5
add  $r5, $r5, $r3
lw   $r6, 0($r5)
add  $r4, $r4, $r6
addi $r2, $r2, 1
slt  $r7, $r2, $r1
j    Loop
Exit:   sw   $r4, 56($r0)
```

The example shows the MIPS pipeline instructions in assembly. The first instruction loads the base address for the key. Second instruction loads the lower word of the key at same register in upper instruction. Third instruction increment the base address of the key. Fourth instruction loads the upper word of the key. Similarly next 8 instructions stored key data to particular registers. Next four instructions are NOP which indicates delay of two clock cycles for key to be loaded. CRYPT 1 instruction enables encryption process. Further, Next instructions are the simple MIPS program instructions whose the output data which are stored in memory location 56. All these MIPS instructions after CRYPT are encrypted if used in encrypted MIPS processor and decrypted if used in decrypted MIPS processor.

## 4. EXPERIMENTAL RESULTS

The complete pipeline processor stages are modelled in VHDL. The syntax of the RTL design is checked using Xilinx tool. For functional verification of the design the MIPS processor is modelled in Hardware descriptive language. The design is verified both at a block level and top level. Test cases for the block level are generated in VHDL by both directed and random way. Result shows the corresponding symbol and an architectural body in the RTL view. For top level verification assembly program are written and the corresponding hex code from the assembler is fed to both RTL design and model the checker module captures and compares the signal from both the model and display the message form mismatching of digital values.

The complete design along with all timing constraints, area utilization and optimization options are described using synthesis report. The design has been synthesized targeting 40nm triple oxide process technology using Xilinx FPGA Virtex-6 (xc6vlx240t-3ff1156) device. The Virtex family is the latest and fastest FPGA which aims to provide up to 15% lower dynamic and static power and 15% improved performance than the previous generation. It is obvious that there is a trade-off between maximum clock frequency and area utilization (number of slices LUT's) because the basic programmable part of FPGA is the slice that contain four LUTs (look up table) and eight Flip flops. Some of the slice can use their LUT's as distributed RAM.

### 4.1 Simulation Results

Different capabilities and features of VHDL lead to various implementation of the design in terms of performance and speed. All the simulation results are based on Xilinx ISE tool [9], using test bench waveform generator. All the individual waveforms of both the encrypted and decrypted MIPS processor simulated using FPGA Virtex-6 device.

#### 4.1.1 Encrypted MIPS ProcessorResult

Fig. 4 shows the default input encrypted instruction memory contents (imemcontents) inside the instruction memory. These encrypted imemcontents values allocate starting address of 0 to 287 memory locations. The default input decrypted data memory contents (dmemcontents) inside the data memory and allocate memory address of 0 to 55 locations. The three 64-bit unencrypted key values stored at starting address of 104 to 151 inside data memory. First two key values is zero and third key value i.e. KIRATPAL (a string of 8 ASCII characters) in terms of Hex value i.e. 0x4b4952415450414c inside the data memory, starting address of 136 to 151 locations. The figure also shows the resultant waveform generated by the 32-bit encrypted MIPS processor. The input is clock of 4ps time period, active high reset which initializes all processor subunit to zero. After clock period 4ps, active low reset all encrypted instructions are loaded and execute. The input is clock of 16232ps, active low reset, and the resultant encrypted output is obtained as cipher data. The output registers values at 2688ps after executed all encrypted instruction from instruction memory. The plain text (input data) is stored at register (4) i.e. 0x00000038 and output cipher data value is stored at memory location address of 56 (decimal). Before 16232ps the memory location is empty as shown in this figure. The resultant value is obtained after 16232ps as shown in above Fig. 8. The lower byte of encrypted cipher data is stored at memory location of 56 (decimal) and upper byte is stored at 63(decimal). Hence, total 64-bit cipher data value is obtained.

Input/plain text – 0x00000038

Key1 – 0x0000000000000000

Key2– 0x0000000000000000

Key3– 0x4b4952415450414c

Output/cipher text – 0x2542b17039a61551

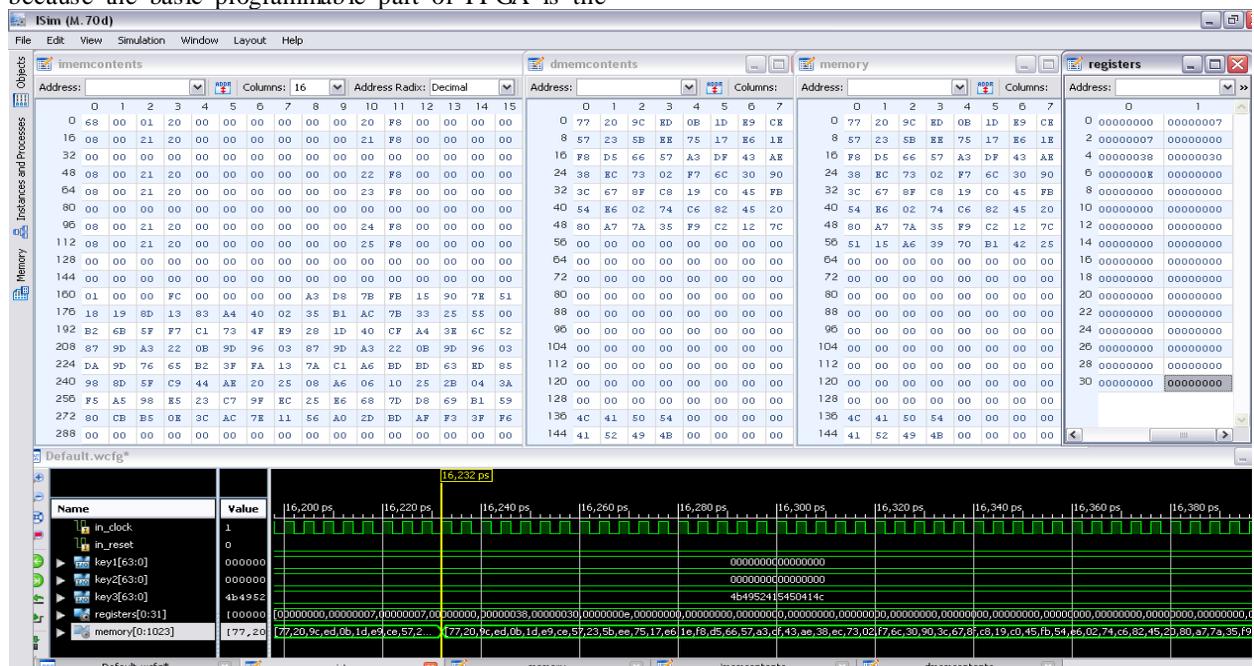

**Figure 3. Input and output values of encrypted MIPS processor and their resultant waveform**

### 4.1.2 Decrypted MIPS processor Result

Fig. 5 shows the default input encrypted instruction memory contents (imemcontents) inside the instruction memory. These encrypted imemcontents values allocate starting address of 0 to 287 memory locations. The default input decrypted data memory contents (dmemcontents) inside the data memory and allocate memory address of 0 to 55 locations. The three 64-bit unencrypted key values stored at starting address of 104 to 151 inside data memory. First two key values is zero and third key value i.e. KIRATPAL (a string of 8 ASCII characters) in terms of Hex value i.e. 0x4b4952415450414c inside the data memory, starting address of 136 to 151 locations. The resultant waveform generated by the 32-bit encrypted MIPS processor. The input is clock of 4ps time period, active high reset which initializes all processor subunit to zero. After clock period 4ps, active low reset all encrypted instructions are loaded and execute. The input is clock of 16232ps, active low reset, and the resultant encrypted output is obtained as cipher data. The output registers values at 16232ps after executed all encrypted instruction from instruction memory. The plain text (input data) stored at register (4) i.e. 0x00000038.The output cipher data value is stored at memory location address of 56 (decimal). Before 16232ps the memory location is empty as shown in this figure. The resultant value is obtained after 16232ps as shown in Figure. The lower byte of encrypted cipher data is stored at memory location of 56 (decimal) and upper byte is stored at 63(decimal). Hence, total 64-bit cipher data value is obtained.The decrypted MIPS processor found the correct value for sum of array i.e. 0x00000038, which is stored in register (4).

Input/cipher text – 0x00000038
Key1 – 0x0000000000000000
Key 2– 0x0000000000000000
Key3 – 0x4b4952415450414c
Output/plain text – 0x2c824fe86704fd6e

### 4.2 Verification and Synthesis

The design is verified both at block level and top level. As system verification, we successfully execute encrypted MIPS program and T-DES encryption and decryption. Test case for block level is generated in VHDL by both directed and random way.

The synthesis and mapping result of encrypted MIPS pipeline processor design are summarized in Table 3. The speed performance of the processor is affected by hardware (i.e. clock rate), instruction set, and compiler. The timing report of encrypted and decrypted MIPS processor shows that our processor works at 4.77ns clock period at synthesis level and 8.116ns clock period at simulation level. The synthesis report shows the area utilization and timing summary. The area utilization is same for both the encrypted and decrypted MIPS processor. Both the processor can work at 123MHz (at simulation level) to fully executed all instructions.

**Table 3. Synthesis report**

| | |
|---|---|
| *Target FPGA Device* | Virtex-6 (XC6vlx240t-3ff1156) |
| *Process Technology* | 40nm |
| *Optimization Goal* | Speed |
| *Max. operating frequency (hardware)* | 209MHz (synthesis level) |
| *Max. operating frequency (software)* | 123MHz (simulation level) |
| *Number of slice registers* | 10411 |
| *Number of slice LUT's* | 69148 |
| *Number of fully used LUT flip flop pairs* | 9305 |
| *Number of bonded IOB's* | 598 |
| *Instruction throughput* | 636Mbits/sec |
| *Latency* | 21 cycles per instruction |
| *Key Length* | 64-bits |
| *Data Length* | 64-bits |

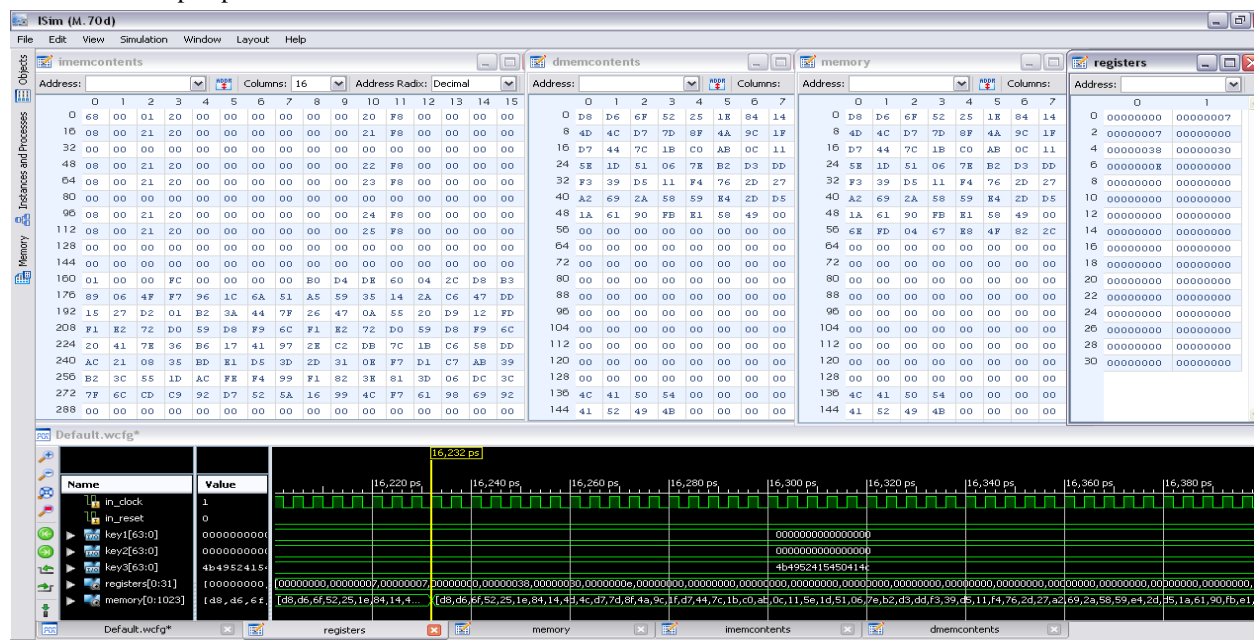

**Figure 4. Input and output values of decrypted MIPS processor and their resultant waveform**

## 5. CONCLUSION

We proposed the high performance 32-bit encrypted and decrypted MIPS processor based on Triple Data Encryption Standard. Which executes encrypted/decrypted instructions, read and decrypt encrypted data from memory unit and write encrypted data back to memory. The processor uses the symmetric block plain/cipher that can process data block of length 64-bits plain text, three 64-bits key and 64-bit cipher data. The design has been modeled in VHDL and functional verification policies adopted for it. Optimization and synthesis of design is carried out at latest and fastest FPGA Viretx-6 device that improves performance. Each program instructions are tested with some of vectors provided by MIPS. The high performance and high flexibility of crypto processor design makes it applicable to various security applications. We conclude that system implementation reach maximum frequency of 209MHz after synthesizing at 40nm process technology and 123MHz at simulation level.